\documentclass[preprint2]{aastex}
\usepackage{amsmath}
\usepackage{mathrsfs}
\usepackage{graphicx}
\usepackage{natbib}
\usepackage[percent]{overpic}
\usepackage[colorlinks=true,citecolor=blue,breaklinks=true,linktocpage=true]{hyperref}
\bibpunct{(}{)}{;}{a}{}{,} 
\usepackage[switch,pagewise]{lineno}
\usepackage{xspace}

\newcommand{\alfven}{Alfv\'en\xspace}
\newcommand{\unit}[1]{\ensuremath{\,\mathrm {#1}}}  
\renewcommand{\d}{\ensuremath{\text{d}}}  
\newcommand{\bvec}[1]{\ensuremath{\boldsymbol{\mathbf{#1}}}}  
\renewcommand{\div}[1]{\nabla\cdot #1} 

\newcommand{\figref}[1]{Figure~\ref{#1}}
\newcommand{\secref}[1]{Section~\ref{#1}}
\renewcommand{\eqref}[1]{Equation~\ref{#1}}
\defcitealias{yuan2015rs}{Paper I} 

\shorttitle{Secondary waves in random plasmas}
\shortauthors{Yuan et al.}

\begin{document}

\title{Secondary fast magnetoacoustic waves trapped in randomly structured plasmas}
\author{Ding Yuan\altaffilmark{1,2}}
\email{DYuan2@uclan.ac.uk}
\author{Bo Li\altaffilmark{3}}
\and
\author{Robert W. Walsh\altaffilmark{1}}
\altaffiltext{1}{Jeremiah Horrocks Institute, University of Central Lancashire, Preston, PR1 2HE, United Kingdom}
\altaffiltext{2}{Key Laboratory of Solar Activity, National Astronomical Observatories, Chinese Academy of Sciences, Beijing, 100012}
\altaffiltext{3}{Shandong Provincial Key Laboratory of Optical Astronomy and Solar-Terrestrial Environment,
Institute of Space Sciences, Shandong University, Weihai, 264209 Shandong, China}

\begin{abstract}
Fast magnetoacoustic wave is an important tool for inferring solar atmospheric parameters. We numerically simulate the propagation of fast wave pulses in randomly structured plasmas mimicking the highly inhomogeneous solar corona. A network of secondary waves is formed by a series of partial reflections and transmissions. These secondary waves exhibit quasi-periodicities in both time and space. Since the temporal and spatial periods are related simply through the fast wave speed, we quantify the properties of secondary waves by examining the dependence of the average temporal period ($\bar{p}$) on the initial pulse width ($w_0$) as well as the density contrast ($\delta_\rho$) and correlation length ($L_c$) that characterize the randomness of the equilibrium density profiles. For small-amplitude pulses, $\delta_\rho$ does not alter $\bar{p}$ significantly. Large-amplitude pulses, on the other hand, enhance the density contrast when $\delta_\rho$ is small but have a smoothing effect when $\delta_\rho$ is sufficiently large. We found that $\bar{p}$ scales linearly with $L_c$ and that the scaling factor is larger for a narrower pulse. However, in terms of the absolute values of $\bar{p}$, broader pulses generate secondary waves with longer periods, and this effect is stronger in random plasmas with shorter correlation lengths. Secondary waves carry the signatures of both the leading wave pulse and background plasma, our study may find applications in MHD seismology by exploiting the secondary waves detected in the dimming regions after CMEs or EUV waves.
\end{abstract}

\keywords{Sun: atmosphere --- Sun: corona --- Sun: oscillations --- magnetohydrodynamics (MHD) --- methods: numerical --- waves}

\section{Introduction}
\label{sec:intro}
Fast magnetohydrodynamic (MHD) waves can propagate across various magnetic structures, and could therefore be easily trapped in structures with low \alfven speeds \citep[see e.g.,][]{goedbloed2004}. Plasma structuring modifies how MHD waves propagate and leads to interesting effects such as wave-guiding, dispersion, mode coupling, resonant absorption, and phase mixing \citep[][]{edwin1983,vandoosselaere2008b,sakurai1991,heyvaerts1983}. Theoretical studies on MHD waves in structured plasmas, combined with the abundant measurements of low-frequency waves and oscillations in the solar atmosphere, can be employed to infer the solar atmospheric parameters that are difficult to measure directly \citep[see the reviews by][and references therein]{nakariakov2005rw,demoortel2012}.
This technique, commonly referred to as solar MHD seismology, has been successful in yielding the magnetic field strength \citep{nakariakov2001}, plasma beta \citep{zhang2015}, transverse structuring \citep{aschwanden2003}, and longitudinal \alfven transit time \citep{arregui2007} in various coronal loops. In addition, it has also been adopted to infer the effective polytropic index of coronal plasmas \citep{vandoorsselaere2011}, magnetic topology of sunspots \citep{yuan2014cf,yuan2014lb},  and the magnetic structure of large-scale coronal streamers \citep{chen2010,chen2011}.

While not a common practice in the literature,
     modelling the inhomogeneous solar atmosphere as randomly structured plasmas 
     is more appropriate in a number of situations. 
For instance, this approach has been adopted to model
     the plethora of thin fibrils in sunspots \citep{keppens1994}, 
     the filamentary coronal loops \citep{parker1988,pascoe2011}, 
     and the structuring in the solar corona across the
     solar disk \citep[][hereafter referred to as \citetalias{yuan2015rs}]{murawski2001,yuan2015rs}. 
\citet{nakariakov2005} examined the dispersive oscillatory wakes
     of fast waves in randomly structured plasmas. 
\citet{murawski2001} studied the possible deceleration of fast waves 
      due to random structuring. 
\citetalias{yuan2015rs} performed a parametric study on the attenuation
     of fast wave pulses propagating across a randomly structured corona,
     and proposed the application of the results for seismologically exploiting 
     the frequently observed large-scale extreme ultraviolet (EUV) waves. 

Previous investigations of the global EUV waves across the solar disk primarily focused on the nature and properties of the leading front \citep[see the reviews by, e.g.,][]{gallapher2011,patsourakos2012,liu2014,warmuth2015}. However, secondary waves have been observed at strong magnetic waveguides or anti-waveguides, e.g., active regions \citep{ofman2002,shen2013}, coronal holes \citep{veronig2006,li2012,olmedo2012}, prominences \citep{okamoto2004,takahashi2015}, and coronal loops \citep{shen2012}. 
These studies on secondary waves were primarily intended to 
    provide support for the wave nature of EUV waves. 
However, given that their spatial distribution and temporal evolution
    can now be observed in substantial detail,
    secondary waves may well be suitable for remotely diagnosing
    the structured solar atmosphere as well (\citetalias{yuan2015rs}).

In this study, we present a detailed numerical study on the interaction between fast wave pulses with a randomly structured plasma, paying special attention to secondary waves in the wake of the leading fast wave pulse. We describe our numerical model in \secref{sec:model}, and then present a case study on the secondary waves and their quasi-periodicity in \secref{sec:period}. Then we perform a parametric study on how this quasi-periodicity is affected by plasma structuring (\secref{sec:para}). Finally, \secref{sec:con} summarizes the present study.
    
\section{Numerical model}
\label{sec:model}
We used MPI-AMRVAC, a finite-volume code \citep{keppens2012,porth2014}, to solve the ideal MHD equations: 
\begin{align}
\frac{\partial\rho}{\partial t}+\div{(\rho\bvec{v})}=&0, \label{eq:mass}\\
\frac{\partial\rho \bvec{v}}{\partial t}+\div\left[\rho\bvec{v}\bvec{v}+\bvec{I}p_\mathrm{tot}-\frac{\bvec{B}\bvec{B}}{\mu_0} \right]=&0, \\
\frac{\partial \epsilon}{\partial t}+\div\left[\bvec{v}(\epsilon+p_\mathrm{tot})-\frac{(\bvec{v}\cdot\bvec{B})\bvec{B}}{\mu_0} \right]=&0, \\
\frac{\partial \bvec{B}}{\partial t}+\div(\bvec{v}\bvec{B}-\bvec{B}\bvec{v})=&0,
\end{align}
where $\rho$ is the density, $\bvec{v}$ the velocity, 
   $\bvec{B}$ the magnetic field, and $\bvec{I}$ is the unit tensor.
In addition, 
   $p_\mathrm{tot}=p+B^2/2\mu_0$ is the total pressure,    
   where $p$ is the gas pressure, and 
   $\mu_0$ is the magnetic permeability of free space.
The total energy density $\epsilon$ is defined by
   $\epsilon=\rho v^2/2+p/(\gamma-1)+B^2/2\mu_0$, 
   where $\gamma$ is the adiabatic index.  

To facilitate the numerical computations, we adopt a set of three independent constants of normalization, 
   namely, $B_0=10\unit{G}$, $L_0=1000\unit{km}$ and $\rho_0=7.978\cdot10^{-13}\unit{kg\,m^{-3}}$.
Some derivative constants are also relevant, 
   e.g., the \alfven speed $V_\mathrm{A}=B_0/\sqrt{\rho_0\mu_0}=1000\unit{km\,s^{-1}}$ 
   and the \alfven transit time $\tau_\mathrm{A}=1\unit{s}$. 
In the following text, all symbols represent normalized variables.

The MPI-AMRVAC code was configured to solve the one-dimensional (1D)
   version of the ideal MHD equations 
   in Cartesian coordinates $(x,y,z)$, meaning that 
   all dependent variables depend only on $y$.
We chose the three-step Runge-Kutta method for time integration.   
Furthermore, from the multiple finite-volume approaches implemented by the MPI-AMRVAC code,    
   we adopted the HLL approximate Riemann solver \citep{harten1983} 
   with the KOREN flux limiter \citep[see, e.g.,][]{toth1996,kuzmin2006}. 

The equilibrium state, into which fast wave pulses are to be introduced,
    is characterized by a uniform, $x$-directed magnetic field
    $\bvec{B}(t=0, y)=(1,0, 0)$. 
The plasma pressure $p(t=0, y)$ is also uniform, corresponding to a 
    plasma beta of $0.01$ everywhere in the computational domain. 
Random structuring is realized by specifying a proper density profile $\rho(t=0, y)$. 
A uniform plasma pressure is maintained by adjusting the temperature profile ($p=\rho T$) accordingly.

The density profile is formed by a set of sinusoidal modulations superimposed on a uniform background,
\begin{equation}
\label{eq:dens}
\rho\left(t=0, y \right) = 1+\frac{\Delta}{N}\sum\limits_{i=1}^{N} R_i\sin \left( \frac{1}{4}\frac{i\pi y}{L_y}  +\phi_i \right),
\end{equation}
where $\Delta$ is a scaling parameter, and $L_y$ is the size of the numerical domain in the $y$-direction. 
The values $R_i$ and $\phi_i$ are the random amplitude and phase of the $i$-th harmonic component, given by uniform pseudo-random number generators within the ranges of $[0,1]$ and [$-\pi$,$\pi$], respectively. 
A correlation length $L_c$ is defined to quantify the average spacing between two fine structures; 
    and a density contrast $\delta_\rho$ is calculated to define the coarseness
    of density fluctuations (\citetalias{yuan2015rs}). 
A density distribution with $\delta_\rho=0.18$ and $L_c=1.26$ 
    is shown in \figref{fig:yt}(a) for illustration purposes.

Fast wave pulses to be launched into the equilibrium are 
    in the form of a Gaussian profile centered around $y_0=0$ 
    with an amplitude of $A_0$ and an initial width of $w_0$ (\citetalias{yuan2015rs}):
\begin{align}
v'_y(t=0, y)&=A_0\exp\left[-4\ln2\frac{(y-y_0)^2}{w_0^2} \right],\\
B'_x(t=0, y)&=A_0\exp\left[-4\ln2\frac{(y-y_0)^2}{w_0^2} \right], \\
\rho'(t=0, y)&=A_0\exp\left[-4\ln2\frac{(y-y_0)^2}{w_0^2} \right],
\end{align}
where $v'_y$, $B'_x$, and $\rho'$ are the perturbations to the $y$-component of the velocity, the $x$-component of the magnetic field, and plasma density, respectively. 
This form of perturbation simulates a local eigenmode solution and ensures that fast waves are uni-directional.

Hereafter, fast wave pulses with $A_0=0.005$ ($0.1$)
    will be referred to as small-(large-) amplitude pulses. 
As will be shown, the small-amplitude pulses have negligible nonlinear effects,
    whereas the large-amplitude ones lead to non-restorable density perturbations.  

\section{Quasi-periodicity of secondary waves}

\label{sec:period}
\begin{figure}[ht]
\centering
\includegraphics[width=0.5\textwidth]{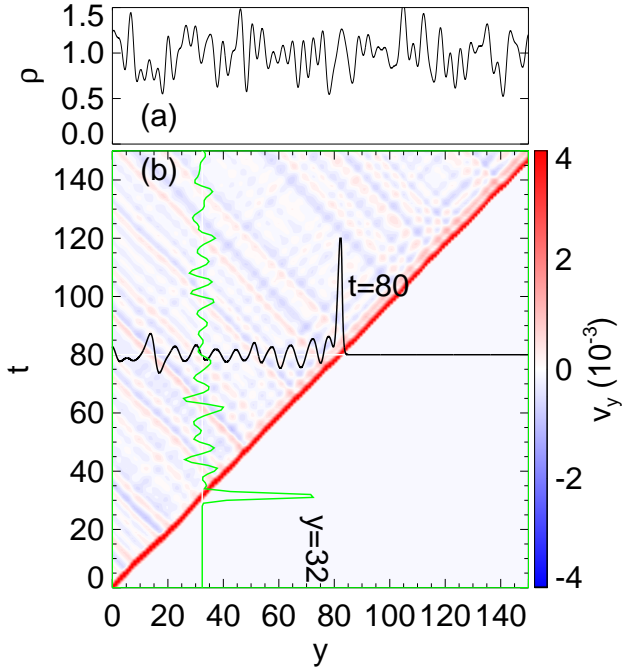}
\caption{
(a) Density distribution of a randomly structured plasma
    with a density contrast $\delta_\rho=0.18$ 
    and a correlation length $L_c=1.26$. 
(b) Distribution in the $y-t$ plane of the $y$-component of the velocity perturbation ($v_y$)
       in response to a small-amplitude fast pulse with $w_0=1.78$. 
    The green (black) curve shows the temporal (spatial) distribution of $v_y$
        at $y=32$ ($t=80$). 
    In both curves, the values of $v_y$ are multiplied by $10^4$ for presentation purposes.
\label{fig:yt}}
\end{figure}

 We start with an examination of what happens when
   a small-amplitude fast pulse with $w_0=1.78$ is launched
   into a randomly structured plasma as depicted in \figref{fig:yt}(a). 
\figref{fig:yt}(b) presents the distribution in the $y-t$ plane of the $y$-component of the velocity perturbation ($v_y$), where the diagonal ridge is the course of the leading fast wave pulse. 
 While propagating, this pulse undergoes a series of interactions with the random plasma, 
   thereby experiencing some attenuation and broadening (\citetalias{yuan2015rs}). 
 In addition, secondary waves with amplitudes at about 10-20\% of that of the main pulse are clearly seen,
   forming a `fabric' pattern as a result of partial transmissions and reflections.
Partial reflection is strong at sharp density changes (e.g., at $y\simeq20, 48, 85, 105$), 
    as evidenced by the strong backward propagation of secondary waves.

\begin{figure}
\begin{overpic}[width=0.48\textwidth]{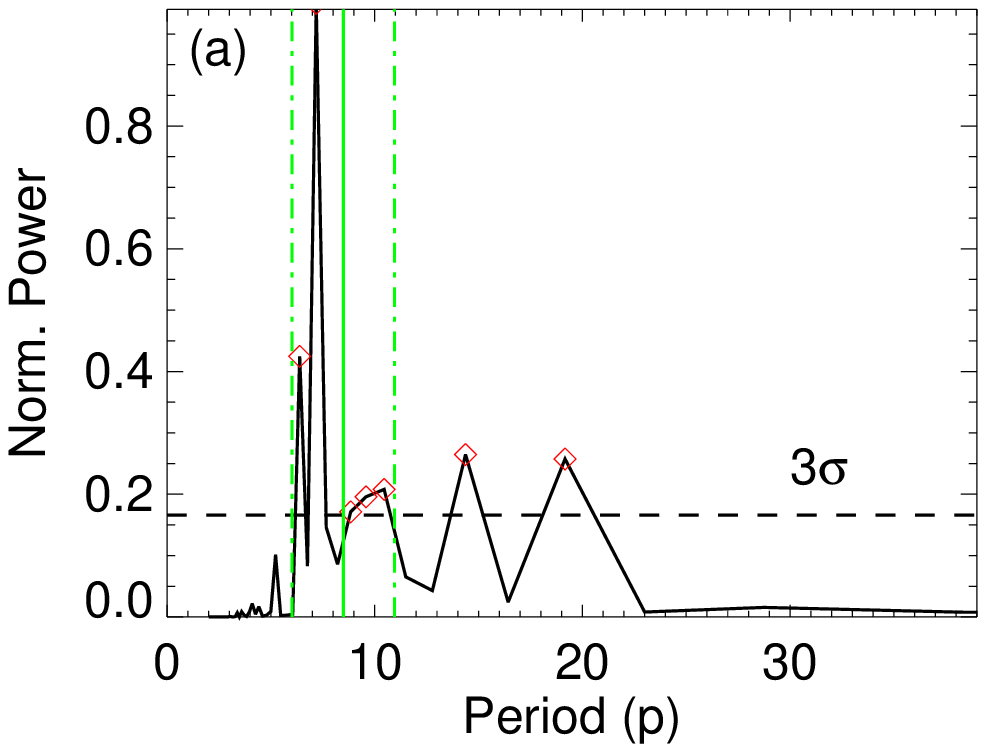}
\put(45,60){\Large\bf $\bar{p}=8.5$}
\put(45,55){\Large\bf $\sigma_p=2.5$}
\end{overpic}
\begin{overpic}[width=0.48\textwidth]{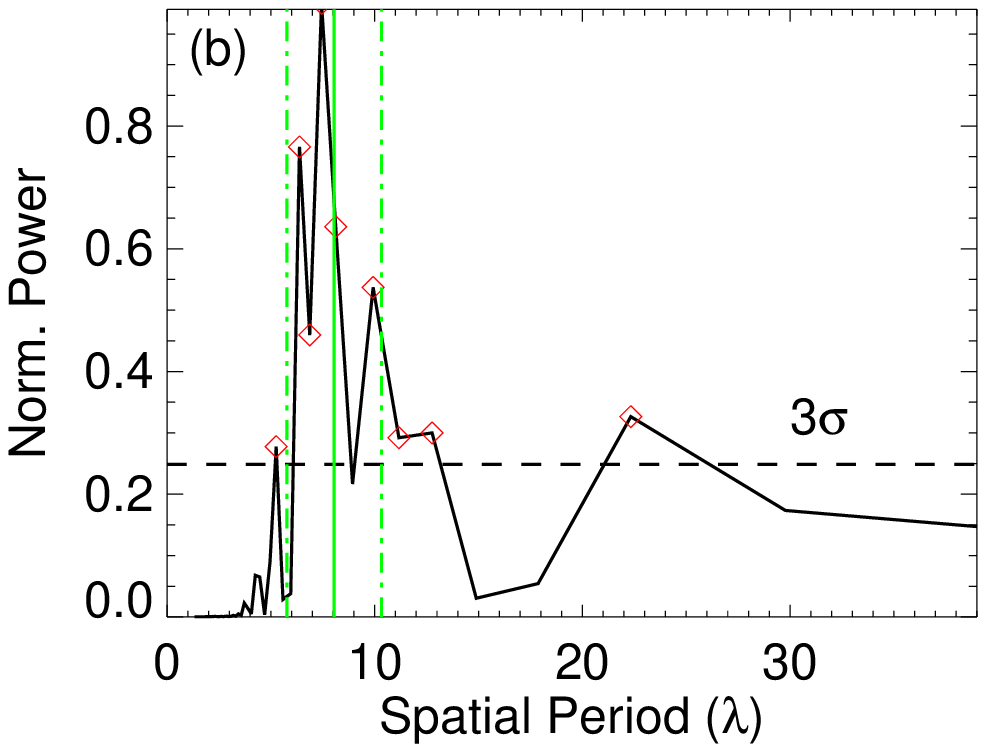}
\put(45,60){\Large\bf $\bar{\lambda}=8.1$}
\put(45,55){\Large\bf $\sigma_\lambda=2.3$}
\end{overpic}
\caption{
Normalized Fourier spectra for (a) $v_y(t, y=32)$ and (b) $v_y(t=80, y)$. 
The solid vertical lines represent the mean periods ($\bar{p}$ and $\bar{\lambda}$), 
    enclosed by the lower and upper limits as given by the dash-dotted lines. 
The horizontal dashed lines plot the $3\sigma$ noise level, 
    and the red diamonds mark the frequency components above it.
\label{fig:fft}}
\end{figure}

We see that the secondary waves exhibit both spatial and temporal quasi-periodicities. 
To better show this, we derive the the Fourier spectra for the secondary waves shown in 
     the two curves in \figref{fig:yt}(b) by excluding the main pulse and the zeros ahead of the pulse.
As shown in \figref{fig:fft}(a), the velocity variation $v_y(t, y=32)$ has prominent oscillations
     with periods in the range between $6$ and $24$ (\figref{fig:fft}(a)).
The average period is found to be $\bar{p}=8.5\pm2.5$ (see Appendix \ref{sec:method} for how we calculate this $\bar{p}$).
This periodicity is not unique to the temporal profile of $v_y$ at $y=32$, 
    but is found at all other positions. 
On the other hand, the spatial period (\figref{fig:fft}(b)) for $v_y(t=80, y)$
    ranges from $6$ to $22$ (or from $5L_c$ to $17L_c$), 
    with an average value of $\bar{\lambda}=8.1\pm2.3$. 
The average spatial period is a few times longer than the correlation length $L_c$,
     meaning that secondary waves need to traverse several correlation lengths
    before settling into a quasi-periodic signal.
This is consistent with \citet{yuan2016st}.
In addition, the temporal periodicity is found to be correlated with the spatial one, which is not surprising 
    given that the fast wave pulse and secondary waves have an average speed of propagation of about unity in numerical units.
The randomness in spatial structuring is transformed into the randomness in the temporal domain.  
As demonstrated in \citet{yuan2016st}, quasi-periodicity is an intrinsic property
   of a random time series, and the quasi-periods are usually a few times
   longer than the timescales of the transients.

\section{Parametric Study}
\label{sec:para}
\begin{figure}[ht]
\centering
\includegraphics[width=0.48\textwidth]{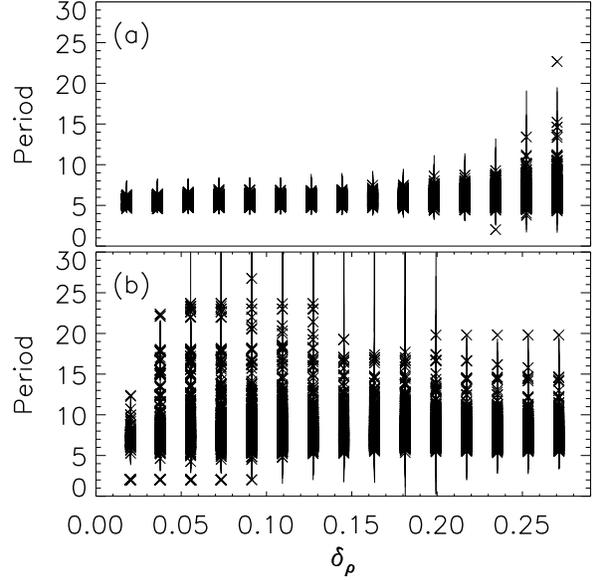}
\caption{
Period as a function of density contrast $\delta_\rho$ for (a) small and (b) large amplitude pulses.
These computations pertain to a fixed correlation length $L_c$ of $1.26$,
    and a fixed initial pulse width $w_0$ of $0.18$.
\label{fig:rho}}
\end{figure}

\begin{figure}[ht]
\centering
\includegraphics[width=0.48\textwidth]{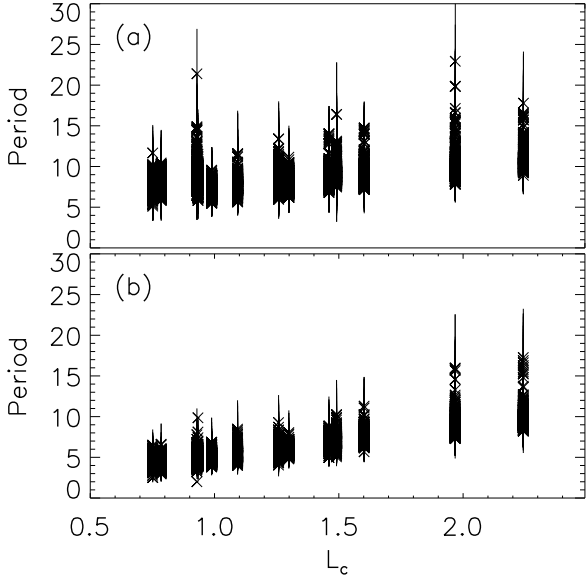}
\caption{
Period as a function of correlation length for small-amplitude pulses
    with two different initial widths: (a) $w_0=2.83$ and (b) $w_0=1.41$. 
These computations pertain to a fixed density contrast $\delta_\rho$ of $0.18$.
\label{fig:Lc}}
\end{figure}

\begin{figure}[ht]
\centering
\includegraphics[width=0.48\textwidth]{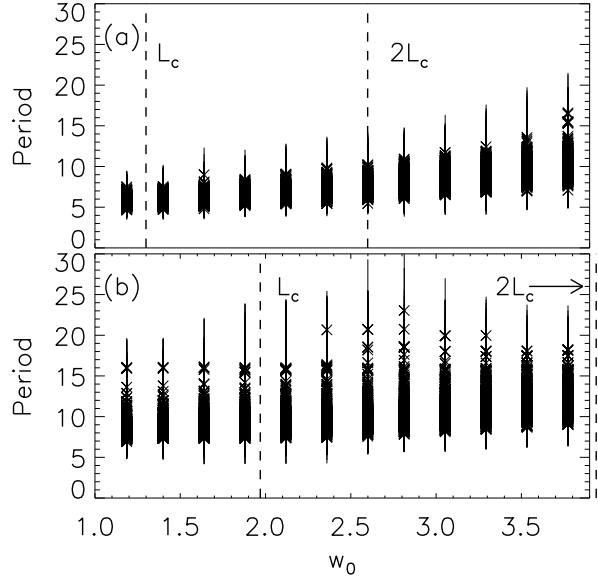}
\caption{Period as a function of initial pulse width. 
Two density profiles  with correlation lengths of (a) $L_c=1.26$  and (b) $L_c=2.0$ are examined,
    even though both pertain to the same density contrast of $\delta_\rho = 0.18$. 
The vertical dashed lines mark where the initial pulse width matches
    $L_c$ and $2L_c$. 
\label{fig:sig}}
\end{figure}

In this section, we investigate how the quasi-periodicities depend on various parameters characterizing
   the equilibrium density profiles and fast pulses.
As is evident from Eq.~(\ref{eq:dens}), the equilibrium density profile is characterized primarily by
    the density contrast $\delta_\rho$ and correlation length $L_c$.
On the other hand, in addition to the amplitudes, fast pulses are also characterized by     
    their initial widths $w_0$.
To bring out the effects of each individual parameter on the average period $\bar{p}$,
    we choose to vary a designated parameter alone by fixing the rest.
In addition, since the periodicities in the temporal and spatial domains are correlated, 
    we will only show the results for the temporal periodicities. 
Given that the time series at each position should be sufficiently long to allow the computation of a significant Fourier spectrum, 
    we choose only the $v_y$ profiles for locations between $y=10$ to $y=100$.
The values for $\bar{p}$ are then scatter-plotted for each parameter, 
    see Figures \ref{fig:rho}, \ref{fig:Lc}, and \ref{fig:sig}. 
A smaller spread of period means that secondary waves are
    trapped by a randomly structured plasma in a more uniform manner, 
    similar to the thermalization of collisional particles heated impulsively.

The dependence of $\bar{p}$ on the density contrast $\delta_\rho$ is shown in \figref{fig:rho},
    for which the values of the correlation length and initial pulse width are fixed
    at $L_c=1.26$ and $w_0=1.18$, respectively.
For small-amplitude pulses, given in \figref{fig:rho}(a),
    the mean period does not vary significantly with the density contrast.
However, the spread in $\bar{p}$ is larger for larger values of $\delta_\rho$.
In the case of large-amplitude pulses, shown in \figref{fig:rho}(b),
    the spread in $\bar{p}$ tends to be stronger than for small-amplitude pulses. 
Furthermore, this spread tends to first increase with increasing $\delta_\rho$ 
    before decreasing when $\delta_\rho \gtrsim 0.2$.
This trend can be understood as follows.    
Large-amplitude pulses can lead to non-restorable density perturbations (\citetalias{yuan2015rs}), 
    and therefore will enhance the density contrast if $\delta_\rho$ is weak. 
However, when $\delta_\rho$ is sufficiently strong, 
    the passage of a nonlinear fast wave pulse has a smoothing effect.

\figref{fig:Lc} shows the dependence of $\bar{p}$ on the correlation length $L_c$. 
In this set of experiments, we fix the density contrast at $\delta_\rho=0.18$ 
   and vary the correlation length. 
Furthermore, while both pertain to small-amplitude pulses, two different values for 
   the initial pulse width ($w_0 = 2.83$ and $1.41$)
   are examined and the results are shown in 
   Figures.~\ref{fig:Lc}(a) and \ref{fig:Lc}(b), respectively.   
It is clear that, regardless of $w_0$,  the mean period $\bar{p}$ tends to depend
   linearly on $L_c$. 
Comparing Figures.~\ref{fig:Lc}(a) and \ref{fig:Lc}(b), one sees that
   a narrow pulse is more sensitive to the variations in the correlation length. 
The period almost triples as $L_c$ doubles.
In contrast, the increase in $\bar{p}$ with $L_c$ for the broader pulse is not as strong. 

How does $\bar{p}$ depend on the initial pulse width $w_0$?
To address this, we launch a set of pulses with different initial widths
    and investigate their propagation in two randomly structured plasmas with
    correlation lengths of $L_c = 2.00$ and $1.26$, respectively. 
The resonant energy loss effect (\citetalias{yuan2015rs}) is not prominent in the periodicity of the secondary waves.
However, \figref{fig:sig} shows that broader pulses normally generate secondary waves with longer periods,
    which is more evident if $L_c$ is smaller.
This is understandable given that when $L_c$ is small,
    fast pulses will be able to interact with the fine-scale inhomogeneities
    more frequently.

\section{Conclusions}
\label{sec:con}

This study offered a series of numerical simulations on the propagation of
    fast MHD wave pulses in randomly structured plasmas mimicking the solar corona. 
While traversing the plasma inhomogeneities,
    fast wave pulses experience partial reflections, giving rise to 
    secondary waves propagating in the opposite direction.
In turn, these waves generate further waves, once again due to their partial reflection
    and transmission at plasma inhomogeneities.
The end result is that, the energy contained in the primary fast pulses
    is spread over the randomly structured plasmas
    in the form of propagating and standing secondary waves. 
These secondary waves exhibit quasi-periodicities in both time and space. 
The spatial period at a given instant is related to
    the period observed at a fixed position via the fast wave speed. 

The interaction between fast wave pulses and the plasma inhomogeneities,
    as quantified by the average temporal period $\bar{p}$ of secondary waves,
    turns out to depend primarily on the combination of parameters
    $[\delta_\rho,  L_c, w_0]$.
Here $\delta_\rho$ and $L_c$ are the density contrast and correlation length that characterize
    the equilibrium density profile.
Furthermore, $w_0$ is the initial width of the wave pulse.
For small-amplitude pulses, $\delta_\rho$ does not have a significant effect on $\bar{p}$.
Rather, it determines the rapidness for $\bar{p}$ to reach a uniform distribution. 
Large-amplitude pulses, on the other hand, lead to non-restorable density perturbations, 
    thereby enhancing the density contrast when $\delta_\rho$ is small but
    having a smoothing effect when $\delta_\rho$ is sufficiently large. 
The average period $\bar{p}$ scales linearly with the correlation length, with the scaling factors 
    being larger for narrower pulses.
However, broader pulses can generate secondary waves with longer periods, 
    the effect being stronger in random plasmas with shorter correlation lengths. 

Secondary waves carrying the signatures of both the leading wave pulse and background plasma may be detected in the dimming regions after CMEs or EUV waves~\citep[see][for some recent observations]{guo2015,chandra2016}. However, a dedicated observational study is needed to fully explore the seismological applications of the present study.

\acknowledgements
We thank the anonymous referee for the constructive comments.
This work is supported by the Open Research Program KLSA201504 of Key Laboratory of
    Solar Activity of National Astronomical Observatories of China (DY). 
It is also supported by the
    National Natural Science Foundation of China (41174154, 41274176, and 41474149).

\appendix
\section{Calculation of quasi-periodicity}
\label{sec:method}
We use $f(s)$ to represent either the spatial distribution
    of the velocity distribution $v_y(y,t)$ at a fixed instant 
    or its temporal variation at a fixed location, 
    after excluding both the leading pulse and the zeros ahead. 
The Fourier transform $\mathscr{F}(\xi)$ is calculated as
\begin{equation}
 \mathscr{F}(\xi)=\int_{-\infty}^{\infty}f(s)e^{-2\pi j s \xi}\d s,
\end{equation}
    where $j = \sqrt{-1}$,
    and $\xi$ denotes either the spatial or temporal frequency.
As a result, $1/\xi$ will be either the spatial or temporal period.
The Fourier spectrum is then obtained by calculating
    $P(\xi)=\left|\mathscr{F}(\xi)\right|^2$. 

The mean frequency $\bar{\xi}$ is taken to be a weighted average,
\begin{equation}
 \bar{\xi}=\frac{\int_{-\infty}^{\infty}P(\xi)\xi\d \xi}{\int_{-\infty}^{\infty}P(\xi)\d \xi}.
\end{equation}
The error bar $\sigma_\xi$ is then computed by using 
\begin{equation}
 \sigma_\xi^2=\frac{\int_{-\infty}^{\infty}P(\xi)\xi^2\d \xi}{\int_{-\infty}^{\infty}P(\xi)\d \xi}-\bar{\xi}^2.
\end{equation}
In our calculations, we integrate only the frequency components
    with power above the $3\sigma$-noise level
    \citep[see][for its definition]{torrence1998}. 
In terms of the temporal or spatial periods, 
    we adopt $1/\bar{\xi}$ and $\sigma_\xi/\bar{\xi}^2$
    as the mean value and its associated uncertainty
    (see the vertical lines in \figref{fig:fft}).

\end{document}